\documentclass[sigconf]{acmart}

\copyrightyear{2026}
\acmYear{2026}
\setcopyright{cc}
\setcctype{by}
\acmConference[SoCC '26 Vol. 2]{ACM Symposium on Cloud Computing Volume 2}{November 18--20, 2026}{Singapore, Singapore}
\acmBooktitle{ACM Symposium on Cloud Computing Volume 2 (SoCC '26 Vol. 2), November 18--20, 2026, Singapore, Singapore}
\acmDOI{10.1145/3828161.3857903}
\acmISBN{979-8-4007-2816-7/2026/11}

\usepackage{booktabs}
\usepackage{multirow}
\usepackage{graphicx}
\usepackage{amsmath}

\graphicspath{{figures/}}

\begin{document}

\title{When, Which, and What to Fork: The Launch-Policy Design Space for LLM Agents}

\author{Zirui Wang}
\affiliation{\institution{Huawei Edinburgh Research Centre}\city{Edinburgh}\country{United Kingdom}}
\email{rwang326@gmail.com}

\author{Mohamed M. Kassem}
\affiliation{\institution{Independent Researcher}\city{Edinburgh}\country{United Kingdom}}
\email{m.m.m.kassem5@gmail.com}

\author{Adam Barker}
\affiliation{\institution{Huawei Edinburgh Research Centre}\city{Edinburgh}\country{United Kingdom}}
\email{adam.barker@huawei.com}

\begin{abstract}
Agents built on LLMs are becoming a major workload for
inference-serving systems. Because a single attempt often fails, operators run
many replicas of the same agent on each task. Each replica, however, is a
stateful stream whose context grows over time. When many such streams share an
engine, they can exhaust its fixed KV-cache memory and hit the memory wall: the
entire batch slows down.

We argue that the missing control layer lies at the client: when to start each
agent and from what state. We propose a launch policy that paces admission by
running leaders first, publishes milestones from their trajectories as
checkpoints, and then releases followers by forking from selected checkpoints.
On SWE-bench Verified, across more than 200 graded runs, we compare our
leader--follower policy with fixed-window launches under the same
live-concurrency cap. On our main task, for batches of 64 agents, even when
fixed-window uses its fastest tested concurrency cap of 26, our policy
reduces batch completion time by $20\%$ and yields $1.36\times$ the solves per
lane-hour, with similar mean resolve rates. For a larger batch of 128 agents
with a concurrency cap of 32, switching to our policy reduces completion time
by $37\%$ and yields $1.7\times$ the solves per lane-hour. These gains come from
the complete policy, and the comparisons do not isolate the contributions of
checkpoint inheritance and selection. Everything runs client-side on an
unmodified engine.
\end{abstract}

\ccsdesc[500]{Computer systems organization~Cloud computing}
\keywords{LLM agents, agent forking, inference serving, admission control, leader selection,
best-of-$N$, KV cache, RL rollout}

\maketitle

\section{Introduction}
\label{sec:intro}
Software agents built on large language models are becoming an important workload for inference-serving systems. Because a single trajectory is often unreliable, practitioners improve success rates by running $N$ replicas of the same agent and keeping the best result, as in best-of-$N$ sampling and reinforcement-learning rollouts~\cite{brown2024monkeys,snell2024scaling,ehrlich2025codemonkeys,shao2024grpo,sheng2024hybridflow}. Although these settings differ in purpose, they produce the same systems workload: many replicas of one coding agent solving one task on a shared inference engine.

Current inference engines schedule \emph{requests}, not \emph{agents}. A coding agent is a long-lived, stateful stream whose context and KV cache grow throughout execution. When many agents share one engine, sequence slots and KV-cache capacity quickly become the bottleneck, causing queueing, preemption, and sharply increasing latency. We call this regime the \emph{KV-cache memory wall}. At the wall, adding replicas no longer improves useful throughput: with the default $K=N/2$ live window, moving from $(N,K)=(64,32)$ to $(128,64)$ reduces fixed-window resolve rate from $54.5\%$ to $24.7\%$, while completion time grows from under 3 hours to roughly 10.

This failure suggests that the system must control not only how requests execute, but also which agents are active and from what state they begin. Launching all $N$ agents immediately from scratch maximizes concurrency and repeats the same early work. Instead, some agents could start later and inherit progress from running peers. This requires deciding \emph{when} to admit each agent, \emph{which} state it should inherit, and \emph{what} state should be reusable. The engine cannot make these decisions because it sees requests rather than agents, while the agent framework runs trajectories independently. We therefore place this control in the \emph{agent runtime}, the client-side orchestration layer between the agent framework and the inference engine, and call the rule governing these decisions the \emph{launch policy}. Given an operator-chosen batch of $N$ attempts, we ask how those attempts should be launched, not how $N$ itself should be chosen.

Existing work addresses either request execution or agent admission, but not stateful agent launch. Within the serving stack, scheduling, paging, disaggregation, and prefix sharing improve how requests execute after they arrive~\cite{kwon2023vllm,yu2022orca,agrawal2024sarathi,qin2025mooncake}. The engine can reuse cached token prefixes, but it cannot restore the modified environment or verified progress associated with an agent trajectory. Agent-level admission control~\cite{chen2026concur} operates at the appropriate granularity but decides only \emph{when} an agent may run; it does not start new agents from other agents' verified checkpoints. The missing capability is a runtime policy that jointly controls admission and state inheritance across agents.

The \emph{launch policy} is an attractive control point because it shapes demand before that demand reaches the inference engine. By controlling how many trajectories are active and whether newly admitted agents repeat or inherit earlier work, it governs both concurrency and redundant computation. It can also be implemented entirely at the client, without modifying the agent, model, or inference engine. We evaluate its system-level benefit using two practical metrics: batch completion time, which captures how quickly all trajectories return, and solves per lane-hour, which captures useful work per unit of accelerator time.

We introduce a leader--follower launch policy. A subset of agents starts first as leaders, each publishing its first checkpoint at the verified reproduction milestone defined in \S\ref{sec:producer}. As leaders finish and free execution slots, followers are admitted from the \emph{deepest} eligible checkpoint, defined as the verified checkpoint with the largest number of completed trajectory steps. Followers therefore inherit the most completed work available instead of repeating the trajectory from the beginning. We use checkpoint depth as a work-reuse heuristic rather than a predictor of eventual success, and compare fork-deepest with fork-random under the same admission rule and checkpoint mechanism to evaluate the selection rule.

Our evaluation includes more than $200$ graded runs comprising about $13{,}000$ trajectories on $19$ SWE-bench Verified instances from $11$ repositories, using Qwen3-Coder-30B-A3B on a commercial ASIC-based AI accelerator. Within each comparison, all policies target the same number \(N\) of attempts under the same per-agent step budget; any unlaunched attempts are counted as unresolved. Any steps inherited by a follower count towards that budget, reducing the number of new steps it may execute. At the capacity-predicted $K=26$ on django-15278 ($N=64$), fork-deepest completes $20\%$ sooner than fixed-window and yields $1.36\times$ the solves per lane-hour. At the KV-cache memory wall ($N=64,K=32$), it reduces batch completion time by $29\%$ and yields $1.5\times$ the solves per lane-hour. Both comparisons hold $N$ and $K$ fixed, so reducing the concurrency cap cannot explain these gains. They evaluate the complete launch policy, combining follower gating, checkpoint selection, and inheritance. Preliminary comparisons without per-task tuning found fork-deepest faster on eleven of thirteen unseen instances.

We make three contributions:

\begin{itemize}

\item \textbf{A launch-policy abstraction for agent workloads.}
We formulate agent launch as a client-side control problem over admission timing, checkpoint selection, and inherited state and cost. This abstraction complements engine-side scheduling and requires no modification to the inference engine.

\item \textbf{A leader--follower policy based on verified work.}
We design an asynchronous checkpoint mechanism through which running agents publish verified progress and later agents resume from it. Deeper checkpoints are associated with less new follower work; timing evidence favoring deepest over random is limited to the tested overload regime.

\item \textbf{A controlled launch-policy evaluation.}
Across more than $200$ graded runs, we compare admission timing,
checkpoint selection, and state reuse under matched budgets. We report
end-to-end improvements in completion time and solves per lane-hour at and
beyond the KV-cache memory wall.

\end{itemize}

\section{Related Work}
\label{sec:related}

\paragraph{Repeated sampling and best-of-$N$.}
Repeated sampling improves success rates in code-generation and agentic workloads~\cite{brown2024monkeys,snell2024scaling,ehrlich2025codemonkeys,antoniades2025swesearch,shum2025swerm}. That line of work treats $N$ as an accuracy knob; we take $N$ as fixed and ask how those $N$ agents should be launched under serving constraints.

\paragraph{Engine-side serving and scheduling.}
Continuous batching, paged attention, request scheduling, and prefill--decode disaggregation optimize requests after they reach the engine~\cite{yu2022orca,kwon2023vllm,agrawal2024sarathi,zhong2024distserve,qin2025mooncake}. Other systems schedule agentic programs using their call structure~\cite{lin2024parrot,luo2025autellix,li2025continuum} or place conversations using observed state~\cite{ding2026conserve}. Our policy acts upstream at the client, deciding when an agent begins and what state it inherits. It is therefore complementary to engine-side scheduling, although we do not measure their composition.

\paragraph{Admission control at the KV-cache wall.}
Admission control limits concurrency to avoid cache
thrashing~\cite{ao2026congestion}. Most closely related,
CONCUR~\cite{chen2026concur} regulates agent concurrency from KV-cache
feedback and pauses and resumes agents without discarding their execution
state. Our policy uses a fixed concurrency cap and launches new followers
from selected peer checkpoints, restoring their transcript and modified
filesystem state. We did not evaluate CONCUR directly because we lacked a
validated integration of its controller and agent-level mechanisms with
our vLLM-based serving stack; reproducing its update rule alone would not establish
a faithful baseline.

\paragraph{KV sharing, forking, and tree search.}
Prefix sharing, forked decoding, and tree-structured search reuse computation within an engine cache or along a single search lineage~\cite{zheng2024sglang,juravsky2024hydragen,gim2024promptcache,yao2023tot,zhou2024lats,wang2026forkexplorecommit}. Our forks operate at agent granularity and include both the transcript and committed filesystem state, allowing them to survive cache eviction and container teardown. We evaluate verified depth as a work-reuse heuristic rather than an outcome predictor (\S\ref{sec:mechanism}).

\paragraph{Rollout scheduling for reinforcement learning.}
Recent RL systems schedule rollouts to reduce straggler cost and, in some cases, bound concurrency against KV-cache capacity~\cite{sheng2024hybridflow,areal2025,seer2025}. Other work branches training rollouts from shared prefixes at token or step granularity~\cite{treepo2025,treegrpo2025,treetraining2025}. Our launch policy brings this idea to serving-time agents by jointly controlling admission, checkpoint selection, and inheritance of the execution environment.

\section{Launch-Policy Design Space}
\label{sec:design}

In our framework, we focus on coding agents and their associated benchmarks, and leave other types of agents for future work. The key idea behind our approach is that a coding agent produces a reusable state that goes beyond its token prefix, i.e., its interaction history, its modified working environment, and its verified progress. Given this richer notion of state, we design a launch policy that coordinates a batch of agents through checkpoints, while leaving the inference engine itself unchanged.

A launch policy must answer three questions: what state to inherit, when a follower may start, and which checkpoint it should inherit. Figure 1 illustrates the resulting pipeline. Leaders publish immutable checkpoints to a shared board, an admission controller bounds the number of live agents at any time, and a selector determines the checkpoint from which each follower resumes. 

\begin{figure}[t]
\centering
\includegraphics[width=\columnwidth]{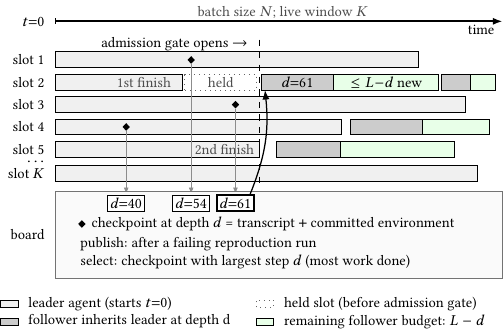}
\caption{\textbf{Client-side leader--follower launch pipeline.}
After the follower-admission gate opens, followers fill available slots by
restoring verified checkpoints. A checkpoint at depth $d$ represents $d$
completed logical agent steps, leaving $L-d$ steps in the follower's total
budget $L$.}
\Description{Timeline diagram of the launch policy. Horizontal bars show K execution slots over time. Leaders start at time zero, and a dashed vertical line marks when the admission gate opens. Diamonds on leader bars mark published checkpoints, which drop onto a checkpoint board holding cards at depths 40, 54 and 61. When a slot frees after the gate opens, a follower starts from the deepest checkpoint (depth 61): its bar has a gray inherited segment followed by a green segment of at most L minus d new steps. A legend explains the leader, held-slot, inherited and new-step segments.}
\label{fig:arch}
\end{figure}

\subsection{What is inherited? Checkpoints and budget}
\label{sec:producer}

One logical step consists of one model query followed by the execution of every
tool action returned in that response. A leader publishes its first checkpoint
at the first logical step $d_1\geq3$ whose actions run a Python program or test
and whose following observation contains a strict failure signature, such as a
traceback, assertion failure, or nonzero failed-test count. We call this event a \emph{verified reproduction milestone}: verified means
only that a failure was observed in execution, not that the target bug was
reproduced or that the patch is correct. The first checkpoint is
the saved artifact of that milestone.

Each checkpoint contains the transcript prefix, its checkpoint depth $d$---the
number of logical steps represented by the saved state---and an
application-level filesystem snapshot, rather than process or inference-engine
KV-cache state. At a logical-step boundary, the runtime serializes the transcript and commits
the leader container's filesystem layer, including repository edits, to a new
Docker image. A follower boots a new container
from that image, loads the transcript as its message history, and sets its
consumed-step count to $d$, so prior tool calls are not re-executed. Its first
model request processes the restored transcript through ordinary prefill.
Automatic prefix caching may reuse blocks still resident from the leader, but
correctness does not depend on a cache hit.

Checkpoints are immutable: work performed by a leader after publication cannot
affect a follower that restores an earlier checkpoint. After publishing its
first checkpoint, the leader continues running. If the same leader later
reaches depth $2d_1$, it publishes a second and final checkpoint from its newer
state. Leaders therefore publish at most two checkpoints, and followers do not
publish, preventing recursive lineages and bounding the checkpoint board.

Finally, the inherited work counts against the normal step budget: an agent may take at most $L$ = 150 steps (i.e., this is based on empirical results on SWE-bench), and a follower that inherits $d$ steps may take at most $L-d$ new steps. Forking therefore never grants an agent more total work than a fresh-start attempt. 

\subsection{When are followers admitted?}
\label{sec:consumer}

The controller targets a batch of \(N\) attempts while maintaining a live window of at most \(K\) agents. It launches \(K\) leaders at \(t = 0\). As agents finish, available slots are refilled one at a time toward the target \(N\), subject to the admission and checkpoint-eligibility rules below. This slot-paced admission bounds simultaneous KV-cache demand and avoids releasing many agents into the engine at once.

For checkpoint-based policies, freed slots are held until the
\emph{follower-admission gate} (hereafter, \emph{admission gate}) opens. Let
$C(t)$ be the number of leaders completed by time $t$. The gate can use either
a \emph{time trigger}, $t\ge c$, or a \emph{leader-completion trigger},
$C(t)\ge m$. The evaluated trigger constants are reported in
\S\ref{sec:eval}.

Fork arms use strict checkpoint admission with no fresh-start fallback. Once
the gate opens, each controller tick admits only followers for which eligible
checkpoints exist and holds any remaining slots. If the board is empty, it
waits while an agent remains active; if none remains, the run ends with the
unlaunched attempts missing, which our metrics count as unresolved.

The main comparisons use $K=N/2$, and checkpoint selectors share the same
admission rule. Section~\ref{sec:window} varies $K$ to identify the live
window that fits the lane's KV-cache capacity.

\subsection{Which checkpoint is selected?}
\label{sec:selector}

When a follower slot becomes available, the selector considers checkpoints
within configured age, context-length, and branch caps. We compare two rules.
\emph{Fork-random} samples an eligible checkpoint uniformly and serves as the
selection control. \emph{Fork-deepest} chooses the eligible checkpoint with
the greatest checkpoint depth $d$, breaking ties in favor of the newer
checkpoint.

Fork-deepest treats depth as a work-reuse score: it chooses the eligible checkpoint containing the largest recorded number of completed logical steps. To isolate
the effect of this choice, we compare fork-deepest with fork-random, which uses
the same pacing and forking mechanism but selects checkpoints randomly.

The evaluation compares selectors under matched admission rules, model,
sampling configuration, hardware, and work budget.

\section{Evaluation}
\label{sec:eval}

\paragraph{Experimental setup.}
We evaluate how to launch a target batch of $N$ agents solving the same software
task. A \emph{run} is one independent execution of one launch policy on one
SWE-bench Verified instance: all $N$ agents share one inference engine, while
$K$ limits how many may be live. Each run exclusively occupies one \emph{lane}:
one vLLM instance with TP=2 on ASIC accelerators. We serve
Qwen3-Coder-30B-A3B-Instruct with vLLM 0.18.0 and the vendor's vLLM plugin.
Engine settings are bfloat16, 131,072-token maximum context, 0.9 memory
utilization, 128-token KV blocks, automatic tool choice with the
\texttt{qwen3\_coder} parser, and prefix caching; chunked prefill is enabled
by the V1 default.

Agents use mini-SWE-agent 2.2.8 at temperature 0.7 and top-$p=1$, with a
6-hour agent timeout and 30-minute model-request timeout. The
controller polls every 5 seconds and enforces a 5-second release cooldown.
Fixed time triggers use $c=900$ and 1,200 seconds for $N=64$ and 128; the thirteen breadth instances (\S\ref{sec:breadth}) use
the leader-completion trigger $m=3$ at $K=32$. An eligible checkpoint is at
most 6 hours old, has at most 65,536 prompt tokens, and has been used by
fewer than two followers. The grader uses SWE-bench 4.1.0. Within each
comparison, we keep the total number of agents $N$, hardware, sampling
configuration, and per-agent step budget $L=150$ fixed across policies.

We run our main replicated experiments on django-15278
(Table~\ref{tab:core}). We also examine operation below the memory wall on
five $N=16$ tasks (django-13012, sympy-16766, sphinx-9711, pytest-10081, and
astropy-14309), comparing one-shot with a fork-deepest variant using
engine-load-based admission, a 30-second cooldown, and fresh-start fallback
($n=4$ per arm). The variant brings no consistent benefit: resolve rates show
no clear difference on any task, and completion time changes by $-6$ to $+5$
minutes.

\paragraph{Metrics and statistics.}
\emph{Resolve rate} is the fraction of the $N$ agents whose submitted patches pass
the SWE-bench grader. \emph{Batch completion time} runs from the first agent
launch until the last agent terminates; engine startup and grading are
excluded. It includes checkpoint publication, follower-container startup, and
transcript prefill; we do not time these components separately. \emph{Solves per lane-hour} divides the number of resolved agents by
batch completion time on one lane. If fewer than the target $N$
attempts are recorded, the missing attempts count as unresolved. Agents in the
same run share an engine queue and cache, so we treat each run---not each
agent---as one independent sample. We compute each metric per run, then report
means and $95\%$ $t$-intervals when $n\ge3$; cells with $n\le2$ are exploratory
and report means only. Direct arm-to-arm comparisons use run-level Welch tests.
Overall, the evaluation contains more than $200$ graded
runs and about $13{,}000$ agent trajectories across $19$ instances.

\subsection{Pacing at and beyond the memory wall}
\label{sec:atwall}

\paragraph{Comparison structure.}
We begin with django-15278, replicated across one-shot, fixed-window,
fork-random, and fork-deepest. One-shot versus fixed-window isolates admission
pacing; fork-random versus fork-deepest isolates checkpoint selection. The
fixed-window versus fork-deepest comparison is the end-to-end policy contrast
and changes both follower gating and inherited state. Unless marked otherwise,
live-window arms use $K=N/2$; the $K=26$ and $K=16$ rows are analyzed in
\S\ref{sec:window}.
The $(N,K)=(64,32)$ and $(128,64)$ fixed-window baselines and fork arms came
from consecutive campaigns with the same model, hardware, budget, window, and
run protocol; no table cell pools runs across campaigns.

\begin{table}[t]
\centering
\fontsize{7.4}{8.6}\selectfont
\setlength{\tabcolsep}{2.4pt}
\renewcommand{\arraystretch}{0.98}
\caption{\textbf{Core results on django-15278.}
Run-level means with $95\%$ $t$-CIs. Resolve rate uses target $N$
(missing attempts are failures). Time is
batch completion time; solves/l-h is computed per run.}
\label{tab:core}
\begin{tabular}{@{}cclcccc@{}}
\toprule
$N$ & $K$ & arm & $n$ & resolve (\%) & time (min) & solves/l-h \\
\midrule
\multirow{8}{*}{$64$}
 & $N$  & one-shot              & 10 & $45.2\pm6.4$  & $222\pm29$ & $8.3\pm2.2$ \\
 & 32   & fixed-window          & 10 & $54.5\pm2.1$  & $146\pm13$ & $14.5\pm1.5$ \\
 & 32   & fork-random           & 10 & $54.7\pm5.6$  & $112\pm19$ & $20.2\pm5.0$ \\
 & 32   & fork-deepest          & 10 & $56.2\pm4.8$  & $103\pm17$ & $22.1\pm4.5$ \\
 & 26   & fixed-window          & 10 & $49.2\pm6.7$  & $91\pm8$   & $21.3\pm3.9$ \\
 & 26   & fork-deepest          & 10 & $53.4\pm3.4$  & $73\pm8$   & $28.8\pm3.7$ \\
 & 16   & fixed-window          & 10 & $54.7\pm5.1$  & $103\pm4$  & $20.4\pm2.5$ \\
 & 16   & fork-deepest          & 10 & $48.8\pm8.6$  & $96\pm5$   & $19.8\pm4.4$ \\
\midrule
\multirow{5}{*}{$128$}
 & 64   & fixed-window          & 3  & $24.7\pm3.0$  & $604\pm83$ & $3.1\pm0.5$ \\
 & 64   & fork-random           & 3  & $27.1\pm24.4$ & $485\pm76$ & $4.2\pm3.2$ \\
 & 64   & fork-deepest          & 4  & $41.0\pm9.9$  & $379\pm97$ & $8.7\pm4.8$ \\
 & 32   & fixed-window          & 10 & $49.2\pm5.8$  & $307\pm27$ & $12.6\pm2.1$ \\
 & 32   & fork-deepest          & 10 & $52.4\pm3.8$  & $193\pm25$ & $21.5\pm3.3$ \\
\bottomrule
\end{tabular}
\end{table}

\begin{figure}[t]
\centering
\includegraphics[width=\columnwidth]{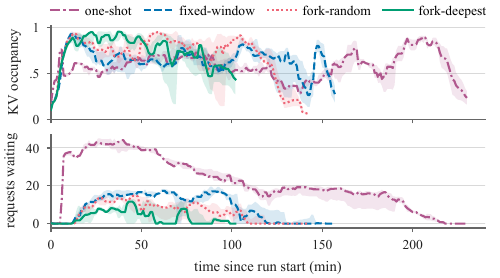}
\caption{\textbf{Engine pressure on django-15278 at $N=64$ (one-shot $K=N$; others $K=32$).}
Lines show per-minute medians across runs; shaded bands show interquartile ranges.}
\Description{Two stacked line charts over time since run start, from 0 to 240 minutes, for four policies at N = 64: one-shot, fixed-window, fork-random and fork-deepest, each drawn with a different line style. The top chart shows KV-cache occupancy between 0 and 1; the bottom chart shows the number of waiting requests. One-shot builds a queue of more than 40 waiting requests that lasts for several hours, while the three live-window policies keep fewer than about 18 requests waiting and finish within about 160 minutes.}
\label{fig:kvcurves}
\end{figure}

\paragraph{Pacing keeps agents out of the engine queue.}
At $N=64$, launching every agent at once achieves a resolve rate of $45.2\pm6.4\%$
and takes $222\pm29$ minutes. Setting the live window to $K=32$, while still
running all $64$ attempts, raises fixed-window resolve rate to $54.5\pm2.1\%$ and
reduces batch completion time to $146\pm13$ minutes. This cuts mean completion
time by $34\%$ and raises solves per lane-hour from
$8.3\pm2.2$ to $14.5\pm1.5$ ($1.7\times$). Figure~\ref{fig:kvcurves} shows why. The one-shot arm peaks above $40$ waiting requests and sustains
queueing for several hours, whereas the live-window policies keep
fewer than approximately $18$ requests waiting.
One-shot nevertheless reports lower average KV occupancy ($0.56$ versus
$0.68$--$0.71$ for the fork arms) because this metric covers only running
requests, not those waiting to enter, and active contexts still need room to
grow. Fixed-window admits work as capacity becomes available, reducing both
the queue and the delay before an agent starts executing.

\paragraph{At the wall, the complete policy improves throughput.}
We next hold the operating point fixed at $N=64,K=32$ and change how later
agents begin. Fixed-window starts each later agent from the original task state;
forked followers inherit the agent context and working environment of a
verified checkpoint. Fork-deepest has a similar mean resolve rate to fixed-window ($56.2\pm4.8$ vs.\
$54.5\pm2.1$), but completes the batch in $103\pm17$ rather than
$146\pm13$ minutes. This $29\%$ reduction in mean completion time raises throughput from
$14.5\pm1.5$ to $22.1\pm4.5$ solves per lane-hour ($1.5\times$). Fork-random
uses the same admission rule and checkpoint format, changing only which
eligible checkpoint is selected; it completes in $112\pm19$ minutes.
All selector-only point estimates favor deepest, including a $9$-minute
($8\%$) completion-time reduction.
With $n=10$ per arm, however, these selector differences remain inconclusive.
We attribute the $29\%$
end-to-end gain over fixed-window to the complete launch policy, which combines
follower gating, checkpoint inheritance, and checkpoint selection. The deepest arm also reaches a higher
prefix-cache hit rate ($0.77$ vs.\ $0.67$ for random). Section~\ref{sec:mechanism} examines how inherited
depth relates to new follower work.

\paragraph{A timing-matched control for the effect of inheritance.}
We replayed the follower-release schedules of five fork-random runs, starting followers from the initial task
state at the recorded release times. Mean completion time was $21$ minutes
shorter with restored checkpoints ($105\pm23$ vs.\ $126\pm38$ minutes; resolve
rates $47.5\pm6.2\%$ vs.\ $55.0\pm5.0\%$), but with five runs per arm this
difference is too uncertain to establish a completion-time benefit. Moreover,
the control replays fork-random, not fork-deepest: random followers inherit a
median of $0.29$ of their leader's trajectory versus $0.54$ under fork-deepest
(\S\ref{sec:mechanism}), so it does not directly evaluate the deeper
inheritance our policy uses. A natural next step is to replay fork-deepest
schedules while also matching the number of live agents over time.

\paragraph{Beyond the wall: an overload stress test.}
The $N=128,K=64$ setting deliberately oversubscribes the engine so that we can
observe how the policies fail under severe contention. Relative to fixed-window,
fork-deepest raises observed resolve rate from $24.7\pm3.0\%$ to $41.0\pm9.9\%$
($1.66\times$), completes the batch in $379\pm97$ rather than $604\pm83$
minutes ($37\%$ sooner), and raises solves per lane-hour from $3.1\pm0.5$ to
$8.7\pm4.8$ ($2.8\times$). This is a stress result, not the recommended operating
point: \S\ref{sec:window} next holds $N=128$ fixed and tests whether reducing
$K$ removes the collapse.
The selector difference is larger under overload. Relative to fork-random,
fork-deepest raises mean resolve rate from $27.1\%$ to $41.0\%$, cuts mean
completion time from $485$ to $379$ minutes ($22\%$), and raises solves per
lane-hour from $4.2$ to $8.7$. With only three fork-random and four fork-deepest runs, we treat the observed 22\% reduction in mean completion time as preliminary evidence favoring deepest selection in this overload setting. Resolve-rate and throughput gains remain variable, and broader generalization requires more selector replications.

\subsection{Sizing the window and gains at a fixed cap}
\label{sec:window}
\label{sec:crossn}
\label{sec:crosstask}

\paragraph{Why the live window has a capacity setpoint.}
Each live agent stores the KV state for its growing context in the engine's
fixed cache. If the lane can hold $B$ KV tokens and a typical live agent uses
$L_{\mathrm{ctx}}$ tokens, then approximately $B/L_{\mathrm{ctx}}$ agents fit
at once. We reserve a fraction $\eta$ of the cache for context growth and
short-term variation, giving the window setpoint
\[
K^\star =
\left\lfloor \frac{B(1-\eta)}{L_{\mathrm{ctx}}} \right\rfloor .
\]
Setting $K$ above $K^\star$ reduces the headroom available for context
growth, increasing the risk of queueing and preemption; setting it far
below $K^\star$ can leave the engine underused. Our lane holds $B=563{,}712$ KV tokens. For django-15278, we set
$L_{\mathrm{ctx}}=17{,}894$ tokens, the median prompt length across model
requests in earlier $N=64$ fork runs. We fixed this estimate and the 15\%
headroom ($\eta=0.15$) before the $K$ sweep; the model then predicts
$K^\star=26$.

\paragraph{A smaller window is not always faster.}
At $N=64$, reducing fixed-window's $K$ from $32$ to $26$ shortens mean
completion time from $146\pm13$ to $91\pm8$ minutes, but a further reduction
to $K=16$ raises it to $103\pm4$ minutes. The window predicted by the capacity
model, $K=26$, is the fastest of the tested settings.

\paragraph{The policy adds value at the chosen window.}
We then hold $N=64,K=26$ fixed and replace fixed-window with fork-deepest.
With $n=10$ independent runs per policy, fork-deepest reduces mean completion
time by $20\%$ ($91\pm8$ to $73\pm8$ minutes) and yields $1.36\times$ the
solves per lane-hour ($21.3\pm3.9$ to $28.8\pm3.7$), while mean resolve rate
is similar ($49.2\pm6.7\%$ vs.\ $53.4\pm3.4\%$). The $K=26$ comparison therefore provides direct evidence of an end-to-end
benefit beyond tuning the fixed-window cap; it evaluates follower gating,
checkpoint selection, and inheritance together. At $K=16$, fork-deepest's completion-time advantage shrinks to $7\%$ and
throughput is nearly unchanged, consistent with underusing the engine.

\paragraph{The collapse follows the live window, not the total batch size.}
At $N=128$, reducing the live window $K$ from $64$ to $32$ cuts mean
completion time from 604 to 307 minutes for fixed-window and from 379 to 193
minutes for fork-deepest, while recovering resolve rate (Table~\ref{tab:core}).
The recovery at fixed batch size identifies live concurrency as the source
of the collapse. At the shared $K=32$, fork-deepest still completes $37\%$
sooner than fixed-window and delivers $1.7\times$ its solves per lane-hour.

\subsection{Inherited depth and follower work}
\label{sec:mechanism}

\paragraph{Followers inherit intermediate progress.}
Let $s$ be the total steps a source leader executes over its whole run, $d$ the
number of steps the follower inherits from that leader's checkpoint, and $F$ the
new steps the follower executes after resuming from that checkpoint.
After the checkpoint, the leader itself executes $s-d$ more steps. A follower whose trajectory is as long as its leader's would therefore
execute $F \approx s-d$ new steps, or, normalized by $s$,
$F/s \approx 1-d/s$.

\begin{figure}[t]
\centering
\includegraphics[width=\columnwidth]{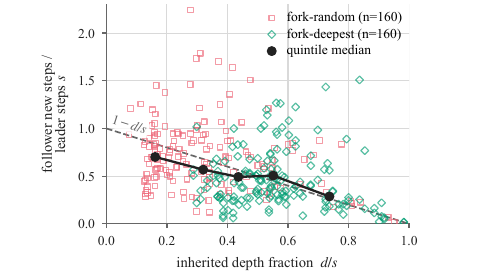}
\caption{\textbf{Deeper checkpoints are associated with less new follower work.}
Two fork arms on django-15278 ($N$=64, $K$=32; five runs per arm, 320
followers). Each open marker is one follower, with inherited $d$ and new
$F$ steps normalized by its leader's total steps $s$. The dashed line
$F/s=1-d/s$ marks the steps the leader itself took after the checkpoint;
followers below it took fewer steps than their leader did from that point.
Filled dots are quintile medians, with both arms pooled and binned by $d/s$.}
\Description{Scatter plot of 320 followers from two fork arms, fork-random and fork-deepest, on django-15278. The x-axis is the inherited depth fraction d/s from 0 to 1; the y-axis is follower new steps divided by leader steps, F/s, from 0 to about 2.3. A dashed reference line F/s = 1 - d/s runs from (0, 1) to (1, 0). Fork-deepest points lie further right than fork-random points, and the quintile medians, shown as black dots, fall along or slightly below the dashed line, decreasing from about 0.7 to 0.3.}
\label{fig:mech}
\end{figure}

Figure~\ref{fig:mech} is consistent with this relationship in both fork arms on
django-15278. Compared with fork-random, fork-deepest has higher median
$d/s$ ($0.54$ vs.\ $0.29$) and lower median $F/s$ ($0.39$ vs.\ $0.68$).
These step-count differences do not establish better batch outcomes;
\S\ref{sec:atwall} evaluates selector performance separately.
Inherited steps also limit followers to $L-d$ new steps, so this trend
does not distinguish avoided repetition from reduced exploration.

\paragraph{Filesystem restoration preserves post-edit state.}
Restoring only a transcript leaves repository edits unrestored. Our checkpoints therefore include both the transcript and the corresponding filesystem state (\S\ref{sec:producer}), allowing followers to resume after repository edits. This capability does not by itself establish how much inheritance contributes to the observed batch-level speedup.

\begin{figure}[t]
\centering
\includegraphics[width=1\columnwidth]{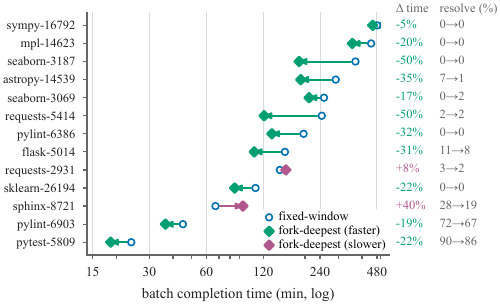}
\caption{\textbf{Preliminary breadth results on 13 unseen instances.}
Mostly one run per arm. Fork-deepest is faster on $11/13$; labels give time change and fixed$\to$fork resolve rate.}
\Description{Dumbbell chart with one row for each of 13 unseen SWE-bench instances, ordered by fixed-window batch completion time on a logarithmic axis from 15 to 480 minutes. Each row connects the fixed-window time, an open circle, to the fork-deepest time, a diamond. Fork-deepest is faster on 11 instances, by 5 to 50 percent, and slower on two: requests-2931 (8 percent) and sphinx-8721 (40 percent). Two columns on the right give each instance's time change and its fixed-window to fork-deepest resolve rates.}
\label{fig:breadth}
\end{figure}

\paragraph{Followers retain room for exploration.}
The selected checkpoint
is a median of $44$ steps before the source leader stops, and every follower
retains at least $64$ of its $150$-step budget. Followers resolve the task both
when their source leaders eventually solve it and when they do not
($53\%$ vs.\ $48\%$). At the
wall, fork-deepest and fixed-window produce $10.2$ and $10.3$ unique resolved
patches per run, and only $17\%$ of fork-deepest followers submit exactly the
patch they inherit. A boundary test starts followers halfway, three-quarters,
or near the end of a leader trajectory; resolve rate falls from $47\%$ to
$28\%$ to $12\%$ as the exploration budget shrinks. Together, these observations
argue against near-terminal solution copying as the explanation for the policy's
gains, but do not rule out reduced trajectory diversity from shared prefixes.

\subsection{Breadth: preliminary cross-task evidence}
\label{sec:breadth}

\paragraph{Transfer without task-specific tuning.}
We evaluate thirteen unseen tasks with matched fixed-window and fork-deepest
runs. Each pair uses $N=64,K=32$ on the same lane with independent cold
starts. Fork-deepest uses the same $m=3$ leader-completion gate across tasks.
Nine have one run per arm; the four with additional replications are
pytest-5809 and scikit-learn-26194 ($n=2$), and astropy-14539 and
requests-2931 ($n=3$).

\paragraph{The policy is faster on most unseen tasks.}
Without per-task tuning, fork-deepest completes $5$--$50\%$ sooner on eleven
tasks (Figure~\ref{fig:breadth}), with fixed-window times from $24$ minutes to 8 hours. Forking has lower
resolve rate on six tasks (by 1--9 points), ties on six, and improves one.
Most cells have $n=1$, and neither policy produced a resolved trajectory on five tasks in the reported runs. These
results provide preliminary evidence that the policy often shortens
completion time, not that it maintains resolve rate. Because the breadth
evaluation has no fork-random arm, it tests the complete policy; fork-deepest is compared with fork-random on replicated django-15278
runs (\S\ref{sec:atwall}).

\section{Discussion}
\label{sec:disc}

\paragraph{Admission pacing helps, but our policy adds to it.}
Admission control is a known remedy for KV-cache
contention~\cite{ao2026congestion,chen2026concur}.
For batches of $N=128$ agents on django-15278, reducing the cap from $K=64$ to
$32$ live agents roughly halves completion time. Our results add two points.
First, our model predicts a cap of $K=26$ live agents from KV capacity and
typical agent context size, and for batches of $N=64$ agents this is also
fixed-window's fastest tested cap on average. Reducing its cap to $K=16$
improves neither mean time nor mean solves per lane-hour.
Second, window tuning alone does not explain the policy's gain: at the same
$K=26$ cap, the complete fork-deepest policy finishes $20\%$ sooner than
fixed-window and yields $1.36\times$ the solves per lane-hour. The
timing-matched control is inconclusive about inheritance's effect on
completion time, and timing evidence favoring deepest over random is
preliminary and limited to the tested overload scenario ($N=128,K=64$).

\paragraph{Forking must save enough work.}
Inherited work pays off only if it outweighs checkpoint wait and restore time
while leaving enough budget for further exploration. In the boundary test,
resolve rate falls from $47\%$ to $12\%$ as followers start nearer a leader's
end, leaving less step budget. A production controller could fall back
to a fixed-window start from the initial task state whenever expected reuse
is too small, a direction we leave for future work.

\paragraph{Scope and limitations.}
We evaluate one model, serving stack, and accelerator platform.
The KV-capacity constraint may transfer, but operating points and gains
are deployment-specific. The replicated matrix concentrates on django-15278,
and the mostly single-run breadth evaluation offers only preliminary evidence
of shorter completion times, not of maintained resolve rates.
CONCUR~\cite{chen2026concur} decides when agents run and preserves each
agent's own state across pauses, but has no cross-agent reuse: it neither
selects peer checkpoints nor restores their transcript and filesystem state.
Without a direct comparison, we claim no performance advantage over it;
composing the two policies, and the effect of shared prefixes on best-of-$N$
diversity, remain future work.
Our checkpoints are also coding-specific, as external tools may have side
effects that cannot be copied safely, and we do not evaluate heterogeneous
or multi-tenant workloads.

\section{Conclusion}
\label{sec:concl}

Inference engines schedule requests, but replicated agents also require
controlling how many trajectories are live and where later trajectories begin.
Our client-side policy bounds live concurrency and restores followers from the deepest eligible checkpoints while charging inherited steps to the same per-agent step budget. On django-15278, lowering the cap from 64 to 32 live
agents removes most of the collapse for a batch of 128 agents. For batches of
64 agents at the capacity-predicted cap of 26, our policy still completes
$20\%$ sooner than fixed-window and yields $1.36\times$ the solves per
lane-hour. In exploratory breadth comparisons without per-task tuning, it completes
sooner on eleven of thirteen unseen instances. Inheritance's separate effect
on completion time remains inconclusive, and evidence favoring deepest over
random selection is preliminary and limited to overload. Admission pacing
helps, and our complete policy adds gains at the same cap on an unmodified
engine.

\clearpage
\balance
\bibliographystyle{ACM-Reference-Format}
\bibliography{references/socc26_camera_ready}

\end{document}